\documentclass[%
 reprint,
superscriptaddress,
 amsmath,amssymb,
 aps,
]{revtex4-2}

\usepackage{graphicx}
\usepackage{dcolumn}
\usepackage{bm}
\usepackage{amsmath}
\usepackage[dvipsnames]{xcolor}
\usepackage[normalem]{ulem}
\usepackage{graphicx}

\newcommand{\Pe}{\textrm{Pe}}
\newcommand{\p}{\textbf{p}}
\newcommand{\I}{\textbf{I}}

\newcommand{\kbT}{k_{\scriptscriptstyle{\textrm{B}}}T}
\newcommand{\br}{\textbf{r}}
\newcommand{\bu}{\textbf{u}}
\newcommand{\bx}{\textbf{x}}

\begin{document}


\title{Hydrodynamic cascade drives tumbling in sheared colloidal rod suspensions}


\author{Lucas H. P. Cunha}
\email{lh1063@georgetown.edu}
\affiliation{Institute for Soft Matter Synthesis and Metrology, Georgetown University, Washington, DC 20057}%

\author{Paul F. Salipante}
\email{paul.salipante@nist.gov}
\affiliation{Polymers and Complex Fluids Group, National Institute of Standards and Technology, Gaithersburg, MD}%

\author{Peter D. Olmsted}
\email{pdo7@georgetown.edu}
\affiliation{Institute for Soft Matter Synthesis and Metrology, Georgetown University, Washington, DC 20057}%

\author{Steven D. Hudson}
\email{steven.hudson@nist.gov}
\affiliation{Polymers and Complex Fluids Group, National Institute of Standards and Technology, Gaithersburg, MD }%


\date{\today}

\begin{abstract}
Modeling the dynamics of colloidal rods remains a central challenge in soft-matter physics due to the anisotropic and long-ranged nature of their interactions. Hydrodynamic interactions in rods suspensions are often assumed to be screened or too week to play any role in semi-dilute regimes, yet we find here these assumptions to break down at shear rates and concentrations that are often attained in experiments. Using particle-based simulations and scaling analysis, we uncover a cascade of tumbling events driven by hydrodynamic coupling among neighboring rods. This collective dynamics disrupts flow alignment and leads to a pronounced increase in viscosity and normal stress differences, in qualitative agreement with recent experiments. The discovery of this hydrodynamically-promoted cascade effect calls for a revision of existing constitutive models for colloidal rods and highlights hydrodynamic coupling as a key mechanism governing collective dynamics in highly anisotropic suspensions.

\end{abstract}

\maketitle


Colloidal rods appear in various biological systems and technological settings, where their orientation distribution dictates the mechanical, thermal, and electrical properties of composite materials. When suspended in a liquid, their elongated shape, Brownian motion, rigidity, and interparticle interactions give rise to intriguing dynamics under flow and complex rheological responses \cite{hinch1972effect, chakrabarti2021signatures, chakrabarti2020flexible, dhont2003viscoelasticity, calabrese2021effects, lang2019microstructural, khan2023rheology, lindstrom2008simulation}. Hydrodynamic interactions (HI) are known to strongly influence the dynamics, gelation, and rheology of colloidal suspensions \cite{turetta2022role, cunha2022, cunha2025hierarchical, varga2015hydrodynamics, furukawa2010key, de2019hydrodynamics, wagner2009shear, jamali2019alternative}, and can even induce self-diffusion in non-Brownian suspensions under flow despite the reversibility of the Stokes problem \cite{morris2011physical, pine2005chaos, leighton1987measurement}.  While HI are known to be key in the rheology of spheres and emulsion droplets \cite{foss2000structure, zinchenko1984effect},  their role in slender bodies remains less understood due to the challenges they pose for analytical and computational modeling \cite{shaqfeh1990hydrodynamic,  butler2018microstructural}.

Under shear flows, the orientation kinetics of a Brownian rod of length $L$ and radius $a$ is dictated by the Péclet number, $\Pe=\dot\gamma/D_r$, where $\dot\gamma$ is the imposed shear-rate and $D_r \simeq  3\kbT \ln(L/4a)/(\pi\eta L^3)$ is the rotational diffusion coefficient \cite{doi1988theory, larson1999structure}. At low-$\Pe$, diffusion dominates and randomizes orientation, while at high-$\Pe$, the rods align with the flow and undergo eventual tumbling \cite{harasim2013direct, schroeder2005characteristic, calabrese2021effects}. In non-dilute conditions, excluded-volume constraints may slow down rotational relaxation depending on the overall orientation, becoming slower at nearly isotropic conditions \cite{kuzuu1983constitutive, doi1981molecular,lang2016connection}. Yet this description neglects HI, whose importance for rod suspensions has long been debated \cite{shaqfeh1990hydrodynamic, batchelor1971stress} and might be relevant especially at high-$\Pe$ \cite{salipante2025rheology}. In water, $\Pe\simeq 0.25(L/\mu\textrm{m})^3\dot{\gamma}(s^{-1})$, so that $\Pe=10^5$ is easily realizable for $L\simeq 10\,\mu\textrm{m}$ rods.

The seminal calculations from \citet{shaqfeh1990hydrodynamic} showed small contributions from HI to the stress in semi-dilute suspensions for rods of large aspect ratio, which has been the basis for neglecting HI in the modeling of rod suspensions \cite{dhont2003viscoelasticity, lang2019microstructural, corona2023testing}. Although this is valid when analyzing the stress in predetermined configurations (\textit{e.g.}, isotropic or aligned), one should still consider HI on the system's kinetics for a complete rheological description, as stressed by \citet{shaqfeh1990hydrodynamic} in the concluding remarks. Recent experiments with fd-virus (an experimental model for Brownian rods) observed high shear viscosities and weak alignment at high-$\Pe$, contradicting previous theories that neglect HI and raising questions about its role \cite{salipante2025rheology}.

\begin{figure*}
    \centering
    \includegraphics[width=1\linewidth]{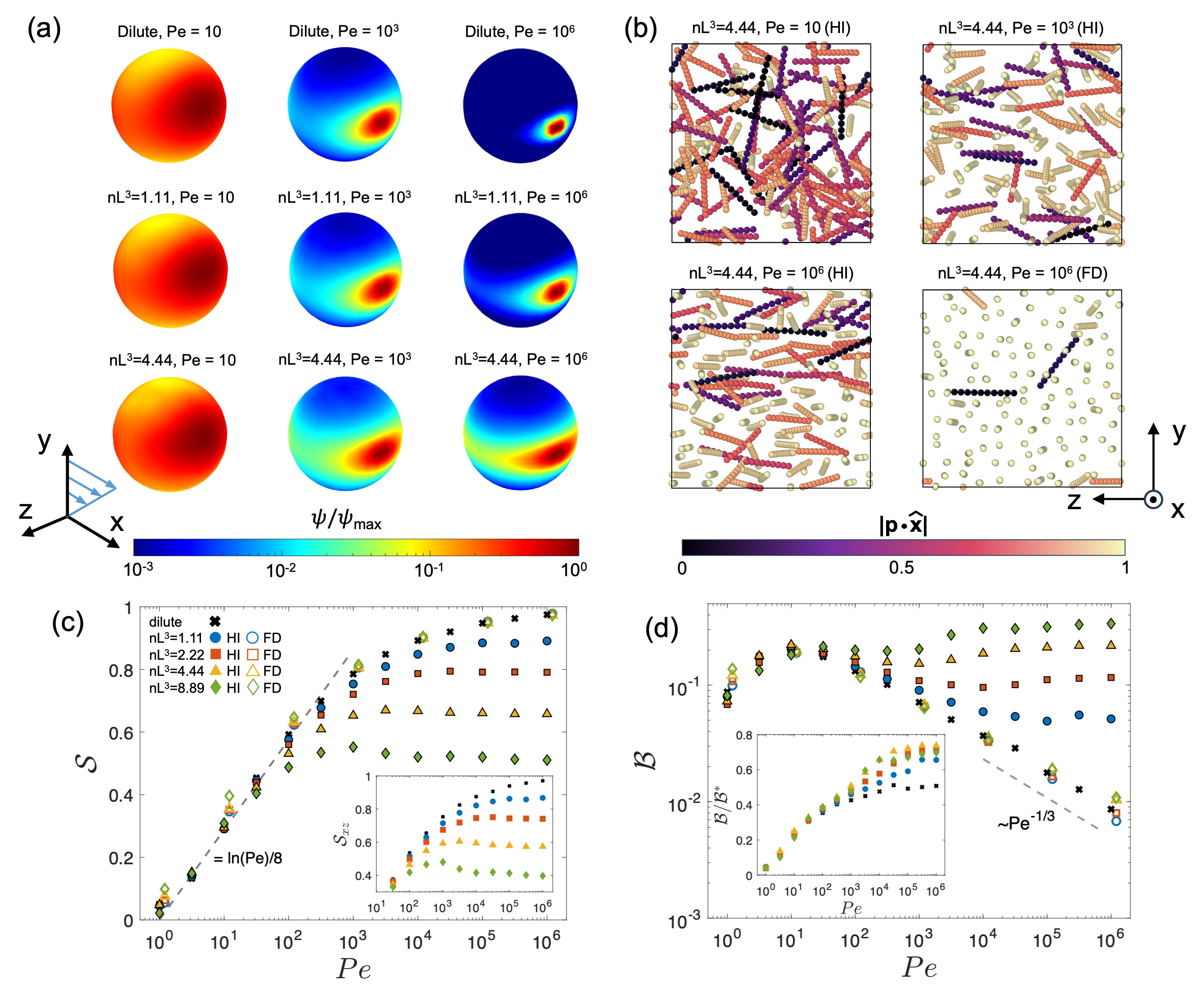}
    \caption{\textbf{Orientation distributions.} (a) Orientation distribution spheres $\psi(\textbf{p})$ for different $\Pe$ (columns) and $nL^3$ (rows). In the orientation spheres, x corresponds to the flow direction, y corresponds to the flow-gradient direction, and z corresponds to the vorticity direction, as shown at the bottom-left. (b) Snapshots of the system configuration viewed in the yz-plane for $nL^3=4.44$ at different $\Pe$, with hydrodynamic interactions (HI) and for the free-draining condition (FD), as informed at the top of each case. The color code represents the absolute value $|\textbf{p}\cdot \hat{\textbf{x}}|$ of the rods' projection in the flow direction.  
    (c) Orientation order $\mathcal{S}$, and (d) biaxiality $\mathcal{B}$ as a function of $\Pe$ for different $nL^3$. Solid symbols represent cases with hydrodynamic interactions (HI), and empty symbols represent free-draining simulations (FD). Inset are the in-plane order parameter $\mathcal{S}_{xz}$ and the biaxiality normalized by its variable maximally attainable value $\mathcal{B}^*$.}
    \label{fig:orientation}
\end{figure*}

To address this, we perform Brownian Dynamics simulations of rods under shear flow, focusing on how HI affects orientation, tumbling, and rheology. We explore a wide range of Péclet numbers ($1\leq \Pe \leq 10^6$) and concentrations in the semi-dilute regime up to $nL^3\leq 8.9$, where $n$ is the number of rods per volume. We find that HI triggers cascades of tumbling events that reduce flow alignment and strongly increase the shear viscosity. Also, HI promotes rod orientation toward the vorticity direction, characterized by the strong increase in the second normal stress difference. Comparisons with `free-draining' FD simulations (\textit{i.e.}, with no HI) show that these concentration-dependent effects arise primarily from hydrodynamic coupling.

The rods consist of $N=10$ torque-free beads of radius $a$, connected by strong bending and stretching constraints, giving a total length $L=2aN$. This model recovers slender-rod dynamics independent of $N$, so dilute non-Brownian rods asymptotically align with the flow instead of tumbling periodically under shear \cite{doi1988theory, jeffery1922motion}. Simulations are performed using Lees--Edwards periodic boundary conditions. To overcome the prohibitive computational cost of HI, we combine a rod-distance-based cut-off criterion with the Rotne--Prager--Yamakawa mobility description \cite{rotne1969, ermak1978brownian}. The full description and simulation parameters are in Methods, while the SI discusses the effects of the hydrodynamic cut-off length.

Figure~\ref{fig:orientation} shows the orientation distribution $\psi$ (Fig.~\ref{fig:orientation}a) and the order parameters $\mathcal{S}$ and $\mathcal{B}$ (Fig.~\ref{fig:orientation}cd) for different $nL^3$ and $\Pe$. Snapshots for $nL^3=4.44$ at different $\Pe$ with HI and under FD conditions are shown in Fig.~\ref{fig:orientation}b. For FD cases, $\Pe$ is defined using $D_r^* = 4\kbT/\pi \eta L^3$ \cite{doi1988theory}. The order parameters are calculated as $\mathcal{S} = (3\nu_1 - 1)/2$ and $ \mathcal{B} = (\nu_2 - \nu_3)/\nu_1$, where $\nu_1\geq \nu_2\geq \nu_3$ are the eigenvalues of the second moment tensor $\textbf{Q} = \langle \textbf{pp} \rangle= \int \psi \textbf{pp} \,\text{d}\textbf{p}$, where $\textbf{p}$ is the orientation vector on the unit sphere. $\mathcal{S}=1$ and $0$ correspond to aligned and isotropic states, respectively, while $\mathcal{B}=1$ and $0$ indicate planar isotropy and uniaxial alignment.

\begin{figure*}
    \centering
    \includegraphics[width=1\linewidth]{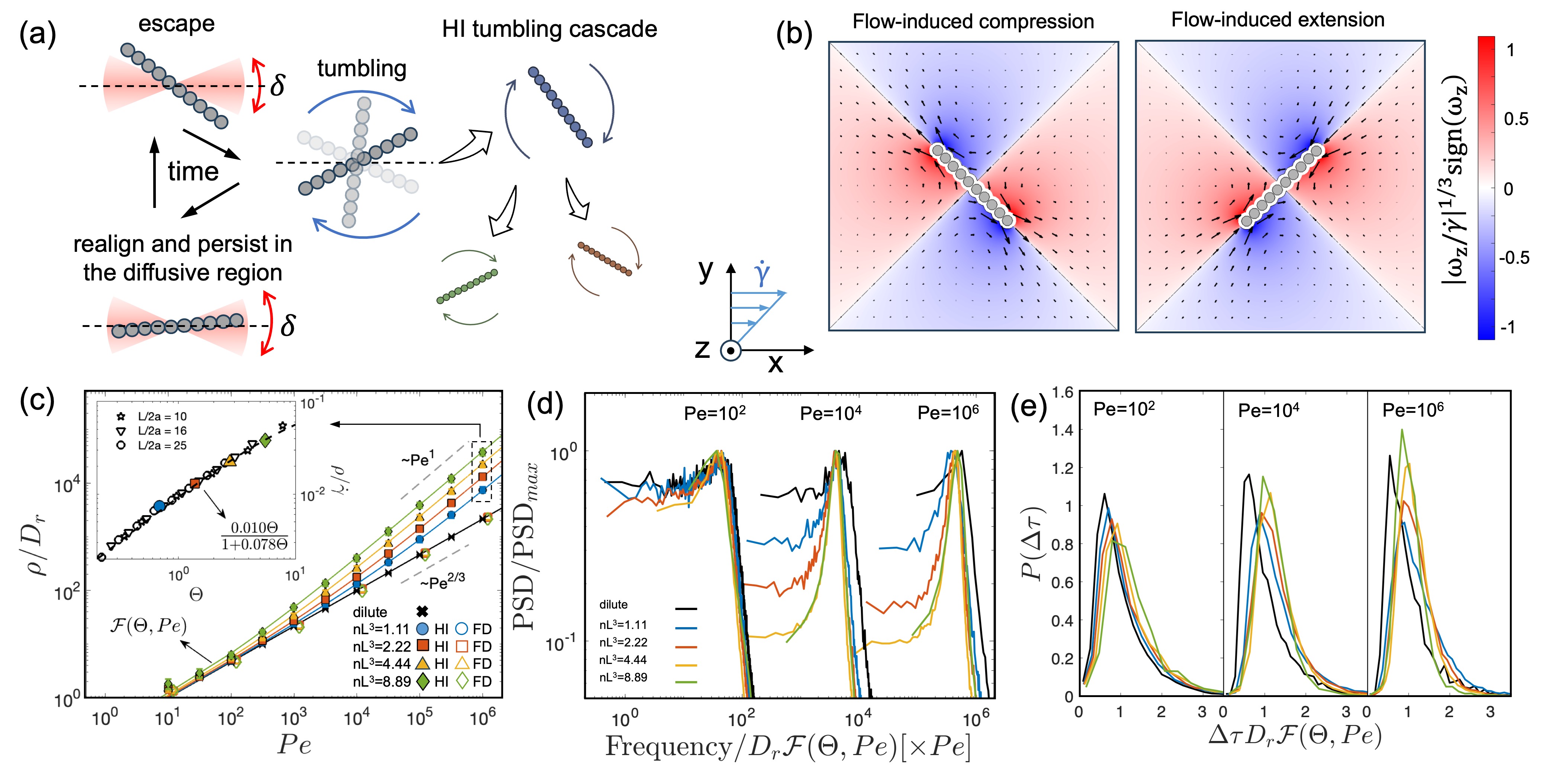}
    \caption{\textbf{Tumbling dynamics.} (a) Sketch of the rod tumbling dynamics and the HI-induced cascade effect, where $\delta$ corresponds to the size of the diffusive dominated region and the empty arrows reflect the promotion of the neighboring rods to also tumble.
    (b) Flow disturbance induced by a rod when oriented in the compressional and extensional directions. The arrows represent the flow field and the color represent the vorticity field scaled as $(\omega_z/\dot{\gamma})^{1/3}$. (c) Mean tumbling frequency normalized by the rotational diffusion $\rho/D_r$ as a function of $\Pe$ for different $nL^3$. Solid symbols represent the HI cases and empty symbols are free draining (FD). The solid lines correspond to Eq.~\ref{eq:tumb_law}.
    The inset presents $\rho/\dot\gamma$ as a function of $\Theta = nL^3/\ln(L/4a)$ for the $\Pe=10^6$ cases (colored symbols in accordance with the main plot), while the black star, nabla, and circle empty symbols represent the results for non-Brownian simulations with HI for $L/2a  =10$, $L/2a  =16$, and $L/2a  =25$, respectively. The dashed black line corresponds to the fitted curve for Eq.~\ref{eq:rho} (d) Normalized power spectral density (PSD) of $x(t)$  as a function of frequency scaled by $D_r\mathcal{F}(\Theta, \Pe)$ for different concentrations and $\Pe$. We multiply each case by the respective $\Pe$ to horizontally separate the curves for different $\Pe$ in the log scale, otherwise all curves fall closely on the top of each other (e) Normalized probability distribution for the modified time lag $\Delta \tau^* = \Delta\tau D_r \mathcal{F}(\Theta,\Pe)$ between consecutive tumbling events for different concentrations.}
    \label{fig:tumbling}
\end{figure*}

At low-$\Pe$, rods weakly align near the flow extensional direction $x=y$ (dark red region in Fig.~\ref{fig:orientation}a for $\Pe=10$). $\mathcal{S}$ increases nearly logarithmically with $\Pe$, independent of $nL^3$, indicating negligible rod interactions at the studied concentrations. $\mathcal{B}$ initially increases with $\Pe$, reaching a peak at $\Pe\sim 10$, followed by a concentration dependent decrease. This biaxiality reflects a weak flow-induced bias of rod fluctuations into the vorticity direction rather than the gradient one. 

For $\Pe\gtrsim 10^2$, the $\mathcal{S}(\Pe)$ and $\mathcal{B}(\Pe)$ curves for different $nL^3$ consistently deviate from one another for the HI cases, while all FD curves fall on top of the dilute one, highlighting the key role of HI. $\mathcal{S}$ decreases with $nL^3$ and plateaus at values $\mathcal{S}(\Pe\rightarrow\infty)<1$.  Interestingly, we capture peaks in $\mathcal{S}(\Pe)$ at $\Pe\approx 2\times 10^3$ for $nL^3 = 4.44$ and at $\Pe\approx 10^3$  for $nL^3 =  8.89$. The peaks are small yet more pronounced at higher concentrations. Analyzing $\mathcal{S}$ projected in the flow-vorticity plane by $\mathcal{S}_{xz} = \sqrt{(Q_{xx} - Q_{zz})^2 - 4Q_{xz}^2}$, we observe more accentuated  peaks (inset in Fig.~\ref{fig:orientation}c). In contrast, the peaks disappear for the projection of $\mathcal{S}$ in the flow-gradient plane, suggesting that they originate from HI biasing the rods' orientations towards the vorticity direction (find the plots for $\mathcal{S}_{xy}(\Pe)$ in the SI). All FD cases show a monotonic decrease in biaxiality  $\mathcal{B}\sim \Pe^{-1/3}$, while for the HI cases $\mathcal{B}$ increases with $nL^3$. 

Increasing $\mathcal{B}$ corresponds to either a smaller $\mathcal{S}$ and/or the greater exploration of $\psi$  in the vorticity direction compared to the gradient one. The later effect may be quantified by $\mathcal{B}/\mathcal{B}^*$, where $\mathcal{B}^*$ is the maximum value of $\mathcal{B}$ for the corresponding $\mathcal{S}$, \textit{i.e.}, $\mathcal{B}(\mathcal{S},\nu_3=0)$.  At high-Pe, $\mathcal{B}/\mathcal{B}^*$ is larger for the HI cases than the dilute one, although the observed response to concentration is non-monotonic (inset of Fig.~\ref{fig:orientation}d). The mechanism behind this non-monotonicity is still unclear.

\begin{figure*}
    \centering
    \includegraphics[width=1.0\linewidth]{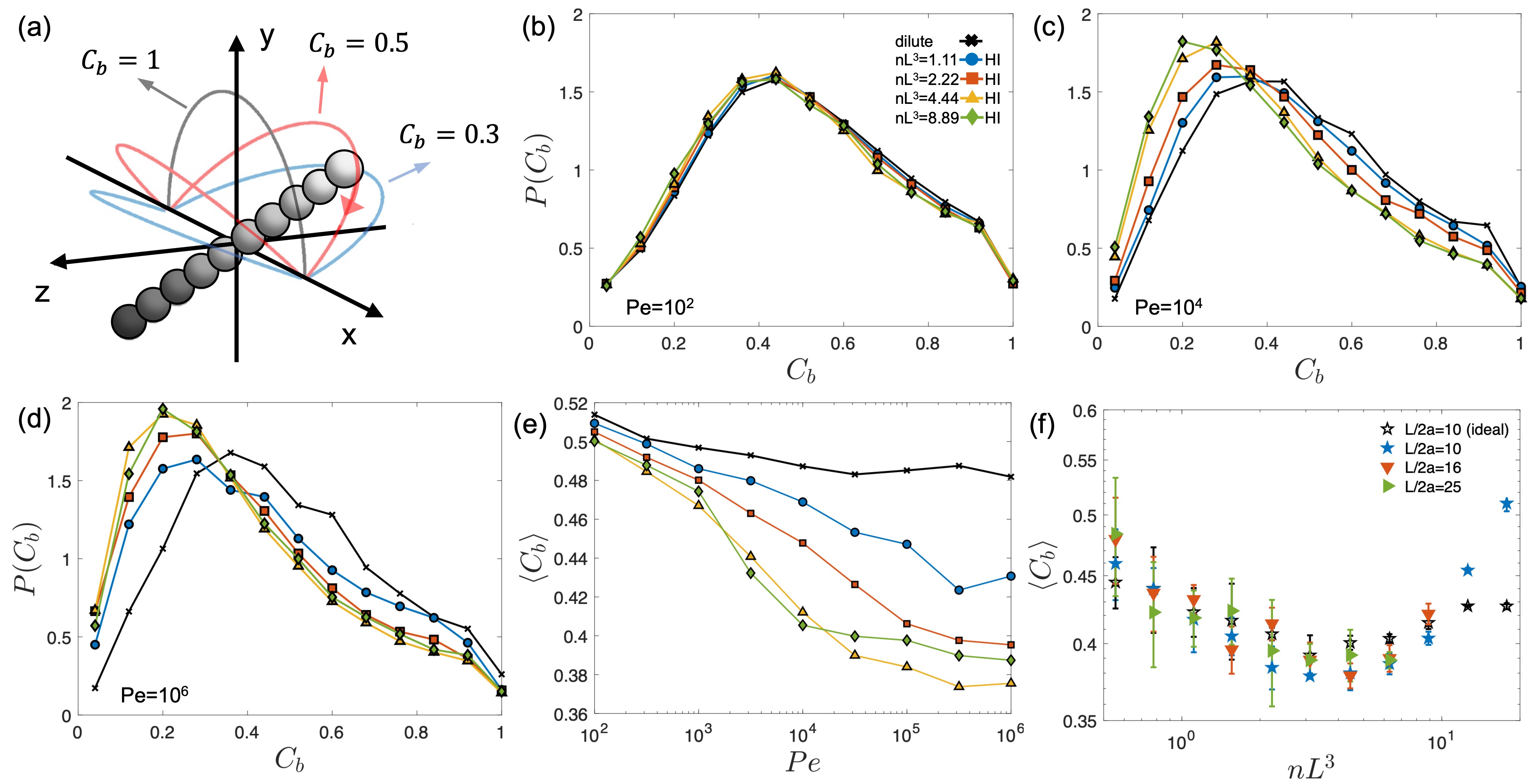}
    \caption{\textbf{Orbit distribution.} (a) Sketch of the different Jeffery orbits for slender rods at different orbit constants ($C_b$). (b)-(d) Distribution of the normalized tumbling orbit constant ($C_b$) for different $nL^3$ at $\Pe = 10^2$, $\Pe = 10^4$ and $\Pe = 10^6$, respectively. (e) Mean tumbling orbit ($\langle C_b\rangle$) as a function of $\Pe$ for different $nL^3$. (f) Mean tumbling orbit as a function of $nL^3$ from non-Brownian simulations for the aspect ratios $L/2a = 10$, $L/2a = 16$, and $L/2a = 25$, and for the ideal case for the aspect ratio $L/2a=10$. }
    \label{fig:orbital}
\end{figure*}

Recent experiments from \citet{salipante2025rheology} for fd-viruses at high-$\Pe$ in pressure-driven capillary flow also showed peaks in $\mathcal{S}(\Pe)$. Due to geometrical constraints, the authors could only access $\mathcal{S}_{xz}$. While the tested fd-viruses have much higher aspect ratio ($L/2a\simeq133$), are more flexible, and the experiments were conducted at different concentrations ($10 \lesssim nL^3 \lesssim 300$), our results suggest that the  experimentally observed peaks are due to HI, since these are not captured by standard models that neglect HI \cite{shaqfeh1990hydrodynamic,lang2019microstructural}.

The tumbling of a slender rod under moderate to high-$\Pe$ is stochastically excited by thermal and/or hydrodynamic fluctuations. First, the flow induces rod alignment until advection is overtaken by diffusion at near alignment. Once the diffusing rod escapes the diffusion-dominated region of size $\delta$ (regions colored in dark red in Fig.~\ref{fig:orientation}a for $\Pe\geq 10^3$) towards the compressional direction ($x=-y$), advection drives the rod to tumble, as sketched in the left side of Fig.~\ref{fig:tumbling}a. \cite{harasim2013direct, hinch1972effect, van2014tumbling}. According to \citet{hinch1972effect}, $\delta \sim \mathrm{Pe}^{-1/3}$ (see SI). While tumbling, the rod is compressed by the flow when in the compressional quadrant, and stretched when in the extensional one. The rods' reaction to the hydrodynamic stresses induces extensional/compressional-like disturbance flows that scale as $ L^3/\ln(L/4a)r^2 $, as shown in Fig.~\ref{fig:tumbling}b (see SI for the description of the flow disturbance). We count a tumbling event when the orientation of a rod crosses from $p_x p_y<0$ to $p_x p_y>0$ between consecutive time-steps.

Previous experiments showed that isolated Brownian filaments tumble at a characteristic frequency $\rho/ D_r \sim \Pe^{2/3}$ \cite{schroeder2005characteristic, harasim2013direct}, as captured by our simulations for the dilute and FD cases when $\Pe \gtrsim 10^2$ in Fig.~\ref{fig:tumbling}c. For the HI cases, this scaling persists up to a critical Péclet number, $\Pe_c$, above which HI overcomes Brownian diffusion and leads to $\rho/D_r \sim \Pe$. Both regimes can be rationalized using simple scaling arguments. For the dilute case, a rod take a period $\tau^d_D \sim \delta^2/D_r$ to diffuse across $\delta$ while the flow takes a period $\tau_A \sim 1/\dot{\gamma}$ to flip them from $\delta$ to $\pi-\delta$, giving $\tau^d_D/\tau_A\sim \Pe^{1/3}$.  Thus, for $\Pe^{1/3}\gg 1$, the tumbling frequency for non-interacting rods satisfies $\rho / D_r \sim 1/\tau^d_D D_r \sim \Pe^{2/3}$. A power fit over the dilute case gives $\rho/D_r = 0.215\,\Pe^{2/3}$.

\begin{figure*}[t]
    \centering
    \includegraphics[width=1\linewidth]{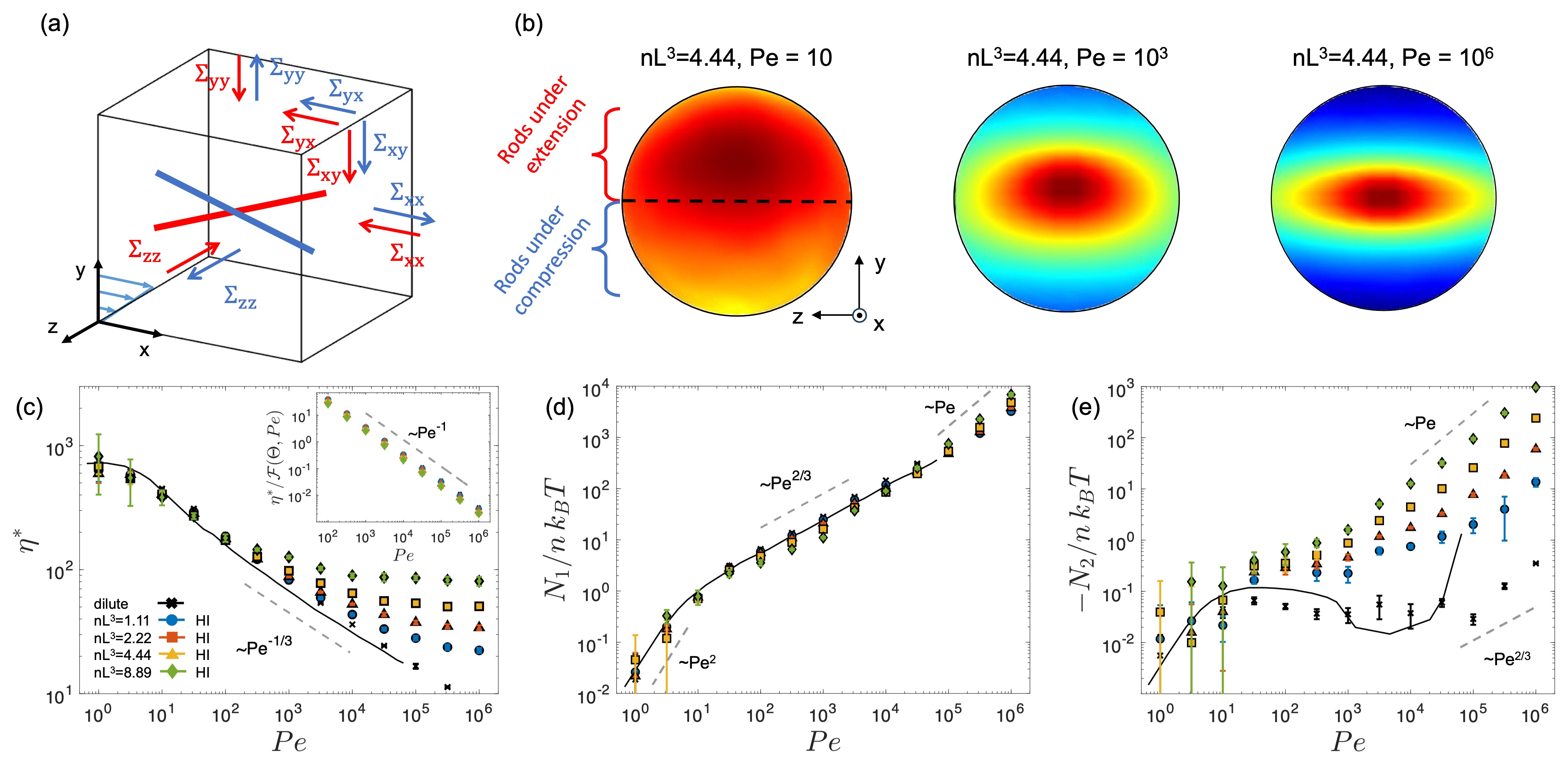}
    \caption{\textbf{Rheological measurements.} (a) Sketch of the stresses induced by the rods when in the extensional (red) and  compressional (blue) quadrants. (b) Orientation distribution $\psi(\textbf{p})$ for different $\Pe=10$, $\Pe=10^3$, and $\Pe=10^6$, and $nL^3=4.44$ in the $yz-$plane perspective. (c) Relative rod viscosity $\eta^* = \Sigma_{xy}/\dot\gamma \eta n a^3$, (d) first normal stress difference $N_1 = \Sigma_{xx}-\Sigma_{yy}$, and (e) second normal stress difference $N_2 = \Sigma_{yy}-\Sigma_{zz}$ as a function of $\Pe$ for different values of $nL^3$. The solid black lines represent the results from \citet{chakrabarti2021signatures}. The error bars correspond to the standard deviation from five to eight independent simulations for the same case.}
    \label{fig:rheo}
\end{figure*}

For semi-dilute suspensions, the effective diffusion should also for hydrodynamic disturbances from tumbling neighboring rods, $D_h$  \cite{rahnama1995effect}. When $\Pe \gg \Pe_c$, hydrodynamics dominates and $D = D_r + D_h \approx D_h$. The HI-induced rotation on a probing rod due to interactions with tumbling neighboring should scale with $\Theta = nL^3/\ln(L/4a)$, see SI. Such interactions result in stochastic kicks at the characteristic frequency $\rho$, implying $D_h \sim \rho \Theta^2$ and modifying the (shear-induced) diffusion time to $\tau^{sd}_D\sim 1/(\dot \gamma^{2/3}/D_h^{1/3})$. Unlike the dilute case, we cannot argue for $\tau^{sd}_D\gg \tau_A$. Instead, we approximate $\rho = 1/(\tau_A + \tau^{sd}_D)$ assuming $\tau_A = A/\dot\gamma$ and $\tau^{sd}_D = B/(\dot\gamma^{2/3}\rho^{1/3}\Theta^{2/3})$, for constants $A$ and $B$. This leads to $\rho A + \rho^{2/3}B\dot\gamma^{1/3}/\Theta^{2/3}=\dot\gamma$, so that we may approximate the solution as
\begin{eqnarray}
\label{eq:rho}
\frac{\rho}{D_r} \approx  \Pe \left(\frac{\Theta/B^{3/2}}{1+ A\Theta/B^{3/2}}\right).
\end{eqnarray}

In the hydrodynamic-dominated regime, $\Pe\gg\Pe_c$, the tumbling frequency of non-dilute Brownian rods should tumble at a frequency similar to their non-Brownian counterpart, as confirmed by the inset in Fig.~\ref{fig:tumbling}c. Fitting the model in Eq.~\ref{eq:rho} to the data for $\Pe =10^6$ gives $A \approx 7.7$ and $B \approx 21.4$.  Combining the dilute and the semi-dilute high-Pe limits,
\begin{eqnarray}
\label{eq:tumb_law}
\rho/D_r \approx \mathcal{F}(\Theta, \Pe) = \left[  \left( 0.215\, \Pe^{2/3}\right)^{m} +\left( \frac{0.010\,\Theta\,\Pe}{1 + 0.078 \Theta } \right)^m \right]^{1/m}, \nonumber \\
\end{eqnarray}
\noindent
valid when $\Pe^{1/3} \gg  1$, where $m\approx3/2$ fits the results the best (solid lines in  Fig.~\ref{fig:tumbling}c). Comparing both terms in Eq.~\ref{eq:tumb_law}, we estimate $\Pe_c \approx [(1+0.078\Theta)/0.047\Theta]^3$. Although our computational model does not recover the tumbling frequency $\rho_J = \dot\gamma r_e /\pi(r_e^2+1)$ predicted by Jeffery equations for dilute rods \cite{jeffery1922motion}, we find that the here described semi-dilute tumbling frequency prevails when $nL^3\gtrsim \rho_J\ln{(L/4a)} /(0.010-0.078\rho_J)$. For cylindrical rods $r_e \approx 1.24(L/2a)/\sqrt{\ln(L/2a)}$ \cite{jeffery1922motion,bretherton1962motion}. Thus, for $L/2a = 10$, 16, and 25, HI dominates, respectively, when $nL^3\gtrsim 8.82$, 6.96, and 5.41, which in the experimental literature represent common conditions \cite{butler2018microstructural, salipante2025rheology, lang2019microstructural, corona2023testing}. 

The tumbling dynamics of Brownian filaments under shear exhibit a single peak in the power spectral density (PSD) of orientation or conformation properties, as shown for $p_x(t)$ in Fig.~\ref{fig:tumbling}d \cite{schroeder2005characteristic}. The peaks appear only for $\Pe\gtrsim10^2$, their frequencies correspond to half of $\rho/D_r$, and collapse for different cases when rescaled by $D_r \mathcal{F}(\Theta, Pe)$. Notably, the peaks become more pronounced with concentration, indicating more regular tumbling intervals $\Delta\tau$ for the same rod. Accordingly, the distributions $P(\Delta \tau^*)$, where $\Delta\tau^*=\Delta\tau D_r \mathcal{F}(\Theta,\Pe)$, narrow at higher concentration for high-Pe (Fig.~\ref{fig:tumbling}e). This reflects the relatively increasing contribution of the advective part of the tumbling dynamics to $\Delta\tau$, consistent with $\tau_A/\tau^{sd}_D = \Theta AB^{-3/2}(1+AB^{-3/2}\Theta)^{-1/3}$.

From Jeffery's description \cite{jeffery1922motion, bretherton1962motion, brenner1974rheology},  an isolated non-Brownian rod undergoes periodic tumbling on closed orbits defined by the constant $C = \sqrt{r_e^2 p_y^2 + p_x^2}/r_e |p_z|$.
For reference, Fig.~\ref{fig:orbital}a shows these orbits for $r_e\rightarrow \infty$ in terms of the rescaled orbit constant $C_b = C/(C+1)$; $C_b = 1$ corresponds to the rod tumbling in the flow plane, and $C_b= 0$ to alignment with the vorticity direction. In Brownian or semi-dilute systems, the rods are not confined to a single orbit but fluctuate between them, resulting in the orbit distribution $P(C_b)$ \cite{butler2018microstructural, rahnama1995effect, hinch1972effect}. Here, $C_b$ is evaluated at the moment at which we capture the tumbling event, and $P(C_b)$ is reported only for $\Pe\geq10^2$, where tumbling is well defined (Fig.~\ref{fig:orbital}b–d).

At $\Pe=10^2$, $P(C_b)$ shows no appreciable dependence on concentration (Fig.~\ref{fig:orbital}b). Increasing $\Pe$ modestly reduces the mean orbit constant in the dilute case, from $\langle C_b\rangle \approx 0.51$ at $\Pe=10^2$ to $\langle C_b\rangle \approx 0.48$ at $\Pe=10^6$ (Fig.~\ref{fig:orbital}e). For $\Pe\gtrsim10^4$, the concentration dependence becomes non-monotonic. Overall, the distributions narrow and the peaks shift to lower $C_b$ relative to the dilute case. At high $\Pe$, increasing $nL^3$ to 4.44 decreases $\langle C_b\rangle$, while further increasing to $nL^3=8.89$ reverses this trend.

The non-monotonic behavior of $\langle C_b\rangle$ with $nL^3$ is also present in the non-Brownian simulations (Fig.~\ref{fig:orbital}f). Increasing $nL^3$ initially decreases $\langle C_b\rangle $ up to $nL^3 \approx 4.5$, followed by an increase at higher concentrations, consistently across all tested aspect ratios ($L/2a=10,16,25$). To better understand the role played by steric interactions,  we run ideal cases for $L/2a=10$ without contact forces, allowing rod overlap. The ideal cases are free of HI singularities due to the Yamakawa regularization in the mobility tensor for overleaping beads (see Methods). The ideal simulations show similar non-monotonic behavior, but the latter increase at $nL^3\gtrsim 4.5$ is less pronounced. This suggests that the initial decrease is led by HI, while the later increase results from combined HI and steric interactions (see SI for analysis on the number of contacts).

We now focus on the rheological response. The rods' contribution to the bulk stress $\mathbf{\Sigma}$ is in great part determined by $\psi$ \cite{doi1988theory, larson1999structure} and is computed here using the inertialess virial formulation, which also accounts for inter-rods interactions (see Methods). Figure~\ref{fig:rheo}a sketches the stress induced by a rod under extension ($p_x p_y>0$, in red) and compression ($p_x p_y<0$, in blue). The larger portion of rods under extension over compression (Fig.~\ref{fig:rheo}b) produces positive first normal stress differences ($N_1 = \Sigma_{xx}-\Sigma_{yy}$) and negative second normal stress differences ($N_2 = \Sigma_{yy}-\Sigma_{zz}$). By symmetry, $\Sigma_{xz}=\Sigma_{yz}=0$.

Our dilute simulations reproduce both the reduced shear viscosity ($\eta^*=\Sigma_{yx}/\dot\gamma\eta n a^3$) and $N_1$ reported by \citet{chakrabarti2021signatures} for a single elastic fiber in the rigid-like regime modeled with slender-body theory (solid black lines in Fig.~\ref{fig:rheo}cd). Agreement is weaker for $N_2$ (Fig.~\ref{fig:rheo}e), although our simulations roughly recover the scaling $N_2\sim \Pe^{2/3}$ theoretically predicted by \citet{hinch1972effect}. At low-$\Pe$, the weak stresses require long simulations for reliable statistics, leading to larger error bars at higher concentrations. Overall, the agreement with \citet{chakrabarti2021signatures} validates our approach for the rheology of Brownian rods.

At low-Pe, we find no appreciable effects of interactions between rods, as characterized by the overlap between all $nL^3$ cases, within error bars. When the system is pushed beyond $\Pe> 10^2$ the rheological signatures of rods interactions become evident. The HI curves for $\eta^*(\Pe)$ consistently deviate from the dilute one, approaching plateau values when $\Pe\rightarrow \infty$. At moderate to high-Pe, the rods' major input to $\eta^*$ occur during the tumbling events. The higher the $nL^3$, the higher the $\rho$ due to the cascade tumbling mechanism, and the higher $\eta^*$. As a consequence, we may collapse  the curves for $\eta^*(\Pe\geq10^2)$ at different concentrations when normalizing by $\mathcal{F}(\Theta,\Pe)$ in Eq.~\ref{eq:tumb_law} (inset of Fig.~\ref{fig:rheo}c). A better collapse is not obtained because such normalization does not account for the shifts in the orbit distributions $P(C_b)$ caused by rods interactions. 

Using Eqs. 30 and 43a in \citet{shaqfeh1990hydrodynamic}, we calculate that the theoretical extra stress from hydrodynamic reflection contribution assuming the isotropic state for $nL^3=8.89$ and $L/2a=10$ corresponds to about 3\% of the total stress and is dominated by the two-body interactions. This confirms that, although the direct contribution from HI to the stress in the system is negligible, its indirect effect from the role played on the system's kinetics can lead to strong rheological consequences. The recent experiments for semi-dilute fd-viruses from \citet{salipante2025rheology} have also captured deviations from the dilute theoretical predictions for the shear viscosity in semi-dilute conditions at high-$\Pe$, suggesting the presence of the cascade tumbling dynamics in their system. The plateaus for $\Pe\rightarrow \infty$ were not experimentally observed, possibly due to shear-induced bending of the fd-viruses at strong shear rates \cite{chakrabarti2021signatures}. 

Regarding $N_1$, we capture a more complex effect from HI (Fig.~\ref{fig:rheo}d). For $10\lesssim \Pe \lesssim 10^4$, increasing $nL^3$ decreases $N_1$, while this relation inverts when $\Pe \gtrsim 10^4$. In the first regime, although there is a greater $D_h$ due to the higher $\rho$ and $nL^3$, the increasing of $\psi$ in the vorticity direction at the cost of the gradient direction results in the weaker $N_1$. For $\Pe\gtrsim 10^4$, the former effect overcomes the latter, increasing $N_1$ with $nL^3$. Lastly, for $N_2$ at high-Pe, we find the strong increase of $-N_2$ with $nL^3$, as a combined response to the HI-induced orientation of the rods towards the vorticity direction and increase in $D_h$.

In summary, we have shown that hydrodynamic interactions reshape the kinetics of semi-dilute suspensions of Brownian rods at experimentally relevant shear rates. Rather than being negligible or screened, hydrodynamic coupling promotes a cascade of tumbling events that disrupts flow alignment and biases orientations toward the vorticity direction. This collective mechanism provides explanations for recent experimental observations of strong hydrodynamic coupling in suspensions of fd-viruses at high-Pe regimes that lead to deviations from classical dilute theories, including enhanced viscosities and the emergence of concentration-dependent features such as peaks in alignment. Importantly, we demonstrate that these rheological signatures arise not from the multi-body hydrodynamic reflection of stresses, but from their indirect impact on the rods’ kinetics, highlighting a subtle yet dominant pathway by which hydrodynamics governs macroscopic behavior.

These findings call for a reassessment of constitutive models for rod-like suspensions, which have largely neglected hydrodynamic interactions in semi-dilute conditions. By establishing scaling laws that capture the transition from Brownian- to hydrodynamic-dominated tumbling, and by connecting microscopic cascade dynamics to bulk rheology, our work provides a framework for incorporating these effects into predictive theories. More broadly, the cascade mechanism identified here suggests that collective hydrodynamic phenomena may play a similarly important role in other highly anisotropic and semi-flexible filamentous systems, from biological to engineered complex fluids.

\bibliography{apssamp}
\section*{Methods}
\textbf{Computational model:} Our computational domain consists of $N_{rods}$ rods in a periodic cubic box of side length $\ell = 3L$, with concentration $n=N_{rods}/\ell^3$. The rods comprise $N=10$ beads of radius $a$, suspended in a Newtonian fluid of viscosity $\eta$ and subjected to a shear rate $\dot{\gamma}$. The centers of neighboring beads on a given rod are connected by massless Hookean springs of relaxed length $2a$, such that $L = 2N a$. The stretching potential for the Hookean spring is $U_s = \kappa_s(r-2a)^2/2$, where $r$ is the center-to-center distance between beads. The filament is made inextensible by imposing a strong stretching rigidity $\kappa_s$. To avoid rod overlap, we implement a repulsive contact interaction between all beads of different rods modeled as a truncated parabolic potential,  $U_c = \kappa_c(r-2a)^2/2$ for $r<2a$, and $U_c = 0$, otherwise. Bending rigidity is accounted for through a parabolic potential $U_b=\kappa_b \theta_b^2/4a$, where $\kappa_b$ is the bending modulus and $\theta_b$ is the angle between neighboring segments \cite{cunha2022}. This potential forces every three neighboring beads to be co-linear, defining the persistence length $\ell_p=\kappa_b / k_B T$. To make the rods stiff, we set a large value for $\kappa_b$. The values used in the simulations for the here mentioned physical parameters are described together with the simulation parameters after the two following subsections.

\textbf{Simulating the dynamics:} The dynamics are given by solving the Langevin equation in the overdamped regime for each bead using the Brownian Dynamics formulation with hydrodynamic interactions (HI) \cite{ermak1978brownian}. The system is made periodic by imposing Lees-Edwards boundary conditions. 

In greater detail, the dynamics of the system follow
\begin{equation}
    \label{eq:BD}
    \mathbf{X}(t+\Delta t) = \mathbf{X}(t) + \mathbb{M}\cdot \mathbf{F^{nh}}\Delta t + \widetilde{\nabla}\cdot(k_BT\mathbb{M})\Delta t + \bm{\Gamma} 
\end{equation}
where $\mathbf{X} = [\textbf{x}_1,\,\textbf{x}_2\,,...,\,\textbf{x}_{n_T}]$ is the beads' position vectors, $N_T = N_{rods} N$, $\widetilde{\nabla}=[\nabla_{\mathbf{x}_1}, \nabla_{\mathbf{x}_2}, \ldots, \nabla_{\mathbf{x}_{N_T}}]$ is the corresponding set of derivatives, $\textbf{F}^{nh} = -[\partial U/\partial \textbf{x}_1,\,\partial U/\partial \textbf{x}_2,\,..., \partial U/\partial \textbf{x}_{N_T}]$ is the non-hydrodynamic forces vector calculated from the total potential $U=U_s+U_b+U_c$, $\Delta t$ is the simulation time-step, $\mathbb{M}$ is the mobility matrix determining the hydrodynamic interactions between all beads, and $\bm\Gamma$ is the Brownian displacement which satisfies the fluctuation dissipation theorem,
\begin{equation}
    \label{eq:tfd}
   \langle \bm{\Gamma}\rangle = \bm0\,\,\,\,\, \text{and} \,\,\,\, \langle \bm{\Gamma}(\Delta t)\bm{\Gamma}(\Delta t) \rangle = 2k_BT\mathbb{M}\Delta t.
\end{equation}
To guarantee that the fluctuation dissipation theorem is obeyed, we implemented the methodology described by Geyer \textit{et al.} \cite{geyer2009n2}. 

\textbf{Hydrodynamic interactions: } In the simulations, we only account for long-range HI and neglect lubrication forces. In the SI, we show that for the concentrations considered in the present study rod collisions are rare
suggesting that lubrication interactions should play only minor effects relative to long-range HI.  The grand  mobility matrix coupling the hydrodynamic interaction between all beads in the system is
\begin{equation}
    \label{eq:mobmat}
    \mathbb{M}
    =
    \begin{bmatrix} 
        \mathcal{M}^*_{11} \,\,\, \mathcal{M}^*_{12} \,\,\, 	\ldots \,\,\, \mathcal{M}^*_{1N_T} \\ 
        \vdots \,\,\,\,\,\,\,\,\,\,\,\,\,\,\,	\ddots   \,\,\,\,\,\,\,\,\,\,\,\,\,\,\, \vdots
        \\  
        \mathcal{M}^*_{N_T 1} \,\,\, \mathcal{M}^*_{N_T 2} \,\,\, 	\ldots \,\,\, \mathcal{M}^*_{N_T N_T} 
    \end{bmatrix} ,
\end{equation}
where the Rotner-Prager-Yamakawa \cite{rotne1969} mobility tensor is used to account for HI:
\begin{widetext}
\begin{equation}\label{eq:rpy}
    \mathcal{M}_{ij} = \begin{cases}
    \dfrac{1}{6\pi \eta a}&i = j \\[14truept]
\dfrac{1}{8\pi\eta}\left[\dfrac{1}{r_{ij}} \left(\mathbf{I}+\mathbf{\hat r}_{ij}\mathbf{\hat r}_{ij} \right) + \dfrac{2a^2}{3r_{ij}^3} \left(\mathbf{I}-3\mathbf{\hat r}_{ij}\mathbf{\hat r}_{ij} \right) \right]  &i \neq j,\,  r > 2a   \\[14truept]
\dfrac{1}{6\pi\eta a}\left[\left(1 - \dfrac{9r}{32a} \right) \textbf{I} + \dfrac{3r}{32a} \mathbf{\hat r}_{ij}\mathbf{\hat r}_{ij} \right]  &i\neq j,\,   r < 2a. 
\end{cases}
\end{equation}
\end{widetext}
Here, $\mathbf{r}_{ij} = \mathbf{x}_{i} - \mathbf{x}_{j}$ and $r_{ij}=|\mathbf{r}_{ij}|$. Note that $\partial \mathcal{M}_{ij}/\partial x_i~=~0$, so that the divergent term (third term in the right hand-side) of Eq.~\ref{eq:BD} vanishes \cite{ermak1978brownian}. Given the $\mathcal{O}(N_T^2)$ computational cost of the simulations when accounting for HI, we implement a rod-based cut-off scheme; otherwise, the required computational time would make the study impractical. For the rod-based hydrodynamic cut-off, we define $\mathcal{M}^*_{ij} = \mathcal{M}_{ij}$ if $i$ and $j$ belong to the same or neighboring rods, and $\mathcal{M}^*_{ij}=\textbf{0}$ otherwise. Two rods are defined as neighbors if their center-to-center distances is smaller than the cut-off distance, here chosen as $L$. The use of the rod-based cut-off scheme instead of a bead-based scheme relies on the fact that all non-hydrodynamic forces in the problem are of an elastic nature and, thus, reciprocal between a pair of beads. In that sense, from the hydrodynamic perspective, one can physically interpret a stretched (or compressed) Hookean springs between beads as a force dipole. The same is valid for the bending forces. Therefore, the hydrodynamic interaction with a bead but not with its pair in the force dipole results in nonphysical long range induced flows from monopoles. This makes the rod-based cut-off scheme an effective solution to implement. An investigation of the choice and effects of the HI cut-off length is presented in the SI. In summary, we study the response of viscosity, orientation distribution, and tumbling frequency to the cut-off distance $r_c$ for non-Brownian simulations setting $L/2a=10$ at the concentrations $nL^3 = 1.11$, $nL^3 = 2.22$, and $nL^3 = 4.44$, considering computational boxes of size $60^3$ and $90^3$. We find that all measurements saturate for $r_c$ greater than $1.5L$. However, using $r_c=1.5L$ results in unpractical simulation times and would prevent the conduction of the present study. Instead, we use $r_c/L = 1$ which led to relative errors in comparison to the $r_c/L=1.5$ cases smaller than $3\%$ for $\mathcal{S}$, $9\%$ for $\rho/\dot\gamma$, and $20\%$ for $\eta^*$. Interestingly, the largest errors for $\rho/\dot\gamma$ and $\eta^*$ do not correspond to the most concentrated case $nL^3=4.44$, but to $nL^3=2.22$. The results suggest that the largest source of error for $r_c = L$  when comparing to $r_c = 1.5L$ is the underestimation of the HI-induced orientation of the rods towards the vorticity direction. More important, we conclude that the artifacts related to cut-off distance do not interfere in any of the conclusions of the present study.

\textbf{Choice of physical and numerical parameters:} The expensive computational cost of the present method and the broad range of $1\leq \Pe\leq 10^6$ considered, requires an adaptable and consistent choice of parameters to reproduce the dynamics of rods. The corresponding high values of $\dot\gamma$ requires equally strong elastic constraints to preserve the rod integrity while requiring finer $\Delta t$ due to their shorter relaxation times. In this sense, the physical and numerical parameters for the simulations were defined as follows: 
\begin{itemize}
    \item $\frac{\kappa_b}{ak_BT} = \textrm{max}\left[250N,\, \frac{4\dot\gamma N^4 \eta a^3}{\log(N/2)k_BT}\right]$
    \item $\frac{\kappa_s a^2}{k_BT} = \textrm{max}\left[0.4\kappa_b a/k_BT,\, \frac{16\dot\gamma N^2 \eta a^3}{\log(N/2)k_BT}\right]$
    \item $\frac{\Delta t \eta a^3}{k_BT}= \textrm{min}\left[2.5\times 10^{-3},\, \frac{\log(N/2)k_BT}{2\dot\gamma N^3 \eta a^3}\right]$
\end{itemize}
On top of that, the simulations account for an adaptive time-step scheme that prevent the largest displacement of a bead in a given direction to be greater than $5\times 10^{-2}a$ to prevent instabilities.

The stress in the system is calculated using the virial formulation in the inertia-free limit,
\begin{equation}
    \label{eq:stress}
    \bm\Sigma = -\frac{1}{\ell^3} \left[ \sum_{ij} \frac{1}{2}(\mathbf{x}_i - \mathbf{x}_j)\mathbf{F}^{ij} \right],
\end{equation} 
where $\mathbf{x}_i$ is the position of  bead \textit{i}, $\mathbf{F}^{ij}$ corresponds to the force applied on bead \textit{i} by bead \textit{j}. This method allows us to capture the rods' contribution to stress due to their orientation distribution as well as the much smaller inter-rod interactions of hydrodynamic or steric natures.

\section*{Disclaimer}
\noindent Certain commercial equipment, instruments, software or materials are identified in this presentation to foster understanding. Such identification does not imply recommendation or endorsement by the National Institute of Standards and Technology, nor does it imply that the materials or equipment identified are necessarily the best available for the purpose.

\section*{Acknowledgments}
\noindent We thank Dr. Mauro Mugnai for insightful discussions. LHPC is grateful for the support of the ISMSM-NIST Postdoctoral Fellowship at Georgetown University. PDO is grateful to the Ives Foundation and Georgetown University for financial support.

\section*{Code and data availability}
\noindent The simulation code used in this study is publicly available at https://github.com/lhpcunha/DynaFibersHI. 
The raw datasets generated during the current study exceed 100 GB in size and are therefore not hosted in a public repository. Processed data supporting the findings of this study, together with selected raw data, are available from the corresponding author upon reasonable request.

\section*{Author contributions}
\noindent All authors contributed to the conceptualization of the investigation. LHPC developed the methodology and implemented the code with feedback from PFS, PDO and SDH. LHPC conducted the simulations and analyzed the data. All authors contributed to the theoretical
framework and interpretation of the results. LHPC wrote the first draft of the paper, with substantial input and revisions from PFS, PDO and SDH. All authors outlined the content of the paper and reviewed and edited the paper.

\section*{Competing interests}
\noindent The authors declare no competing interests.

\section*{Additional information}
\noindent\textbf{Supplementary Information} is available for this work. \\
\noindent\textbf{Correspondence and requests for materials} should be addressed to LHPC or SDH.

\clearpage
\onecolumngrid
\section{Supplemental Information}

\subsection{Theory}
\label{sec:theory}
\subsubsection{Orientation and rheology coupling}

The contribution of rods to the rheological response of suspensions in concentrations far from the nematic transition arises, in one part, from their responses to the compressional and extensional stresses induced by the flow, and in another part, from the entropic losses due to the flow-induced alignment. The first leads to dissipative stresses and the latter to elastic ones, and are respectively given by \cite{doi1988theory}, 
\begin{equation}
    \label{eq:v_stress}
    \bm\Sigma^V = \frac{n k_BT}{2 D_r}\left[\langle\mathbf{pppp}\rangle - \frac{1}{3} \mathbf{I}\langle\mathbf{pp}\rangle  \right]:\mathbf{E}
\end{equation}
and
\begin{equation}
    \label{eq:e_stress}
    \bm\Sigma^E = 3n\kbT\left[\langle \mathbf{pp}\rangle - \frac{1}{3}\mathbf{I}  \right],
\end{equation}
where $\mathbf{E} = (\nabla \textbf{u} + \nabla \textbf{u}^T)/2 $ is the strain rate component of the flow, $\textbf{u}$ is the imposed flow, $n$ is the number of rods per volume, $D_r$ is the rotational diffusion of the rods, $k_BT$ is the thermal energy, $\textbf{I}$ is the unity tensor, \textbf{p} is the rod orientation, and the angle brackets $\langle \cdot \rangle$ refer to the average over a large sample of rods contained in an element of fluid. The rotational diffusion of a rod reads \cite{doi1988theory,tirado1984comparison}
\begin{equation}
    \label{eq:Dr}
    D_r = \frac{3k_BT[\ln(L/4a)+\sigma]}{\pi\eta L^3}
\end{equation}
\noindent
where $L = 2aN$ is the length of the rod, $a$ is the radius of the rod, $N$ is the number of beads used for the rod discretization, $\eta$ is the viscosity of the solvent, and $\sigma$ is a correction term for end effects and rod geometry, e.g., cylindrical or spherical (see \citet{tirado1984comparison, batchelor1970slender, mackaplow1996numerical} for more details). Important to reinforce that Eqs.~\ref{eq:v_stress} and ~\ref{eq:e_stress} do not account for inter-rods interactions of any nature. For that, more elaborated approaches are required. Here, we refer the reader to the works of \citet{shaqfeh1990hydrodynamic} for rods hydrodynamic interactions and \citet{lang2019microstructural} for rods contact interactions effects.

The orientation dynamics of a slender non-Brownian rod in an imposed linear flow is given by \cite{doi1988theory}
\begin{equation}
    \label{eq:rod-dynamics}
    \dot{\textbf{p}} = \nabla \textbf{u} \cdot\textbf{p}\cdot(\textbf{I} - \textbf{pp}).
\end{equation}
On top of the deterministic flow-driven dynamics, colloidal rods are subjected to thermal fluctuations due to their small sizes, resulting in Brownian rotation characterized by $D_r$. Therefore, the orientation distribution $\psi$ for an isolated rod, or equivalently a system of non-interacting rods, is governed by the Fokker-Plank equation
\begin{equation}
    \label{eq:fokker-plank}
    \frac{\partial \psi}{\partial \tilde{t}} + \Pe\frac{\partial }{\partial \textbf{p}} \cdot (\psi \tilde{\dot{\textbf{p}}}) = \frac{\partial }{\partial \textbf{p}}\cdot \left(\frac{\partial \psi}{\partial \textbf{p}} \right),
\end{equation}
where $\partial/\partial \textbf{p}$ corresponds to the derivative on the surface of the unit sphere, time is non-dimensionalized by the rotational diffusivity $t = \tilde{t}/D_r$, $\dot{\textbf{p}} = \tilde{\dot{\textbf{p}}}\dot\gamma$, and 
\begin{equation}
    \label{eq:Pe}
    \Pe=\frac{\dot\gamma}{D_r}
\end{equation}
\noindent
is the P\'eclet number. To account for steric interactions between the rods, which play a considerable role in concentrated systems, we refer to the Doi-Edwards-Kuzuu theory \cite{kuzuu1983constitutive, doi1978dynamics} or later theories that they inspired \cite{lang2019microstructural}. 

There is no analytical expression for the solution to Eq.~\ref{eq:fokker-plank}, although some limits can be explored \cite{hinch1972effect}. If one is specifically interested in the rheological response of the system, we note from Eqs.~\ref{eq:v_stress} and~\ref{eq:e_stress} that the rod contribution to the stress does not explicitly depend on $\psi$, but on the second and fourth moments, $\textbf{Q} = \int \psi \textbf{pp} \, d\textbf{p}$ and $\textbf{Q}^{(4)} = \int \psi \textbf{pppp} \, d\textbf{p}$ \cite{kuzuu1983constitutive}. In this spirit, one may operate on both sides of Eq.~\ref{eq:fokker-plank} with $\int (\textbf{pp} - \textbf{I}/3 )\, d\textbf{p}$, as demonstrated by \citet{doi1981molecular}, to obtain
\begin{equation}
    \label{eq:first-moment}
    \frac{\partial \textbf{Q}} {\partial \tilde{t}}  = -6 \left ( \textbf{Q} - \textbf{I}/3 \right ) + \Pe\left (\nabla \textbf{u} \cdot \textbf{Q}+ \textbf{Q}\cdot\nabla \textbf{u} ^T - 2 \textbf{Q}^{(4)}:\nabla \textbf{u} \right),
\end{equation}
The term $\textbf{Q}^{(4)}$ cannot be directly calculated from $\textbf{Q}$, but is governed by an evolution equation similar to Eq.~\ref{eq:first-moment} with an explicit dependence on the sixth order moment $\textbf{Q}^{(6)}$; and so on for the dynamics of higher order moments. To overcome this issue, different authors assume different closure relations for $\textbf{Q}^{(4)} = \textbf{Q}^{(4)}(\textbf{S})$; for a comprehensive discussion of closure relations see \citet{corona2023testing}.

\subsubsection{Disturbance flow from tumbling rods and the hydrodynamic diffusion}
Here, we calculate the flow disturbance induced by a rod based on the \textit{shish-kebab} model described by Doi \cite{doi1988theory}. The rod is discretized as a linear array of $N$ touching beads of radius $a$, so the total length is $L=2aN$. Following Doi \cite{doi1988theory}, when subject to linear flow $\textbf{u}$, the non-hydrodynamic force acting on bead $i \in (-N/2, N/2)$ reads,
\begin{equation}
    \label{eq:fi}
    \textbf{f}_{i} = -\frac{8\pi\eta a^2}{\ln(N/2)}i \nabla \mathbf{u}:\textbf{ppp}.
\end{equation}
From the Oseen formulation, the disturbance flow at a position $\textbf{r}$ relative to the rod's center of mass induced by the forces (or tension) in the rod is calculated as
\begin{equation}
    \label{eq:dist-u}
    \textbf{u}'(\textbf{r}) = \sum_i \frac{\textbf{f}_i}{8\pi\eta} \cdot \mathcal{G}(\bm{x}_i) = - \frac{a^2 \nabla \mathbf{u}:\textbf{ppp}}{\ln(N/2)} \cdot \sum_i i \left(\frac{\textbf{I} + \hat{\bm x}_i \hat{\bm x}_i }{x_i} \right ), 
\end{equation}
where $\mathcal{G}(\bm{x}_i) = (\textbf{I} + \bm{\hat x}_i\bm{\hat x}_i)/x_i$ is the Oseen tensor, $\bm x_i = \textbf{r} - 2ai\textbf{p}$, $x_i = |\bm{x}_i|$, and $\hat{\bm{x}}_i = \bm{x}_i/x_i$. To approximate the sum on the right-hand-side of Eq.~\ref{eq:dist-u}, we Taylor expand the Oseen tensor around the rod's center of mass, 
\begin{equation}
    \label{eq:oseen-taylor}
    \mathcal{G}(\bm x_i) \approx \mathcal{G}(\bm r) - 2ai\textbf{p}\cdot \nabla \mathcal{G}(\textbf{r}) + \frac{(2ai)^2}{2}\textbf{pp}:\nabla \nabla \mathcal{G}(\textbf{r}) + \mathcal{O}\left ((ai)^3/r^4 \right ),
\end{equation}
and use it in Eq.~\ref{eq:dist-u} to find \footnote{$\nabla_i \mathcal{G}_{jk} = -x_i\delta_{jk}/x^3 + \delta_{ij}x_k/x^3 + \delta_{ik}x_j/x^3 - 3x_ix_jx_k/x^5$}
\begin{eqnarray}
    \label{eq:dist-u-approx2}
    \textbf{u}'(\textbf{r}) \approx  \frac{a^3 N^3 \nabla\textbf{u}:\textbf{pp}}{6\ln{(N/2)}} \left[ \frac{\textbf{r}}{r^3} - 3\frac{(\textbf{p}\cdot  \textbf{r})^2 \textbf{r}}{r^5} \right ]
\end{eqnarray}
\noindent 
for large $N$. Equation is a valid expression in the far-field limit due to the chosen truncation for the expansion in Eq.~\ref{eq:oseen-taylor}. Note that only the terms with odd exponents $i^{2n+1}$ in Eq.~\ref{eq:oseen-taylor} lead to non-zero contributions in Eq.~\ref{eq:dist-u} due to the symmetry around $i=0$ in the sum operator in Eq.~\ref{eq:dist-u}. 

Now, we analyze the the rotation of the rod $\alpha$ with orientation $\p_\alpha$ due to the flow induced by neighboring rods $\beta$ of orientation $\p_\beta$, where $\beta\neq\alpha$ and using $\br_{\beta\alpha}$ as the rods' center-to-center distance. The rotation of $\p_\alpha$ reads
\begin{equation}
    \label{eq:dotp1}
    \dot{\textbf{p}}_\alpha = \nabla \textbf{u} \cdot\textbf{p}_\alpha\cdot(\textbf{I} - \p_\alpha\p_\alpha) +  \sum_{\beta\neq\alpha }\dot\p_{\beta\alpha}'.
\end{equation}
\noindent
The first term on the right-hand-side of Eq.~\ref{eq:dotp1} corresponds to the rotation of $\p_\alpha$ due to the imposed flow while the second term corresponds to the superposition of the rotation due to flow induced by the many neighboring rods. This superposition is  only allowed due to the linearity of the Stokes regime. We may calculate $\dot{\textbf{p}}'_{\beta\alpha}$ as \cite{rahnama1995effect},
\begin{equation}
    \label{eq:dotpab}
    \dot{\textbf{p}}'_{\beta\alpha} = \frac{3}{a^2 N^3}\sum_{i} 2ai(\I-\p_\alpha\p_\alpha)\cdot \bu'_{\beta}(\bx_i).
\end{equation}
\noindent
where $\bu'_{\beta}(\bx_i)$ corresponds to the flow induced by the rod $\beta$ at the position $\bx_i = \br_{\beta\alpha}+2ai$ of bead $i$ belonging to rod $\alpha$. To calculate Eq.~\ref{eq:dotpab}, we use an expansion of $\bu_\beta$ around the center of center of rod $\alpha$, such that
\begin{eqnarray}
    \label{eq:u-exp}
    \frac{\textbf{u}_\beta'(\bx_i)}{c} \approx \mathcal{J}_{\beta\alpha} + 2ai\p_\alpha\cdot\nabla \mathcal{J}_{\beta\alpha} + \frac{(2ai)^2}{2} \p_\alpha\p_\alpha:\nabla\nabla\mathcal{J}_{\beta\alpha} 
\end{eqnarray}
\noindent 
where $c = a^3 N^3 \nabla\textbf{u}:\textbf{pp}/6\ln{(N/2)}$ and $\mathcal{J}_{\beta\alpha} = \br_{\beta\alpha}/r_{\beta\alpha}^3 - 3(\p_\beta\cdot\br_{\beta\alpha})^2\br_{\beta\alpha}/r^5_{\beta\alpha}$. Using Eq.~\ref{eq:u-exp} in Eq.~\ref{eq:dotpab}, we get to \footnote{$\partial_i \mathcal{J}_{j} = \delta_{ij}/r^3 - 3r_j r_i/r^5  - 6(\p\cdot\br)p_i r_j/r^{5} - 3(\p\cdot\br)^2 \delta_{ij}/r^{5} + 15(\p\cdot\br)^2 r_j r_i/r^{7} $}

\begin{equation}
    \label{eq:dotpab-3}
    \dot{\textbf{p}}'_{\beta\alpha} \approx \frac{(a N)^3 \nabla\textbf{u}:\p_\beta \p_\beta}{6\ln{(N/2)}} (\I-\p_\alpha\p_\alpha)\cdot (\p_\alpha \cdot \nabla \mathcal{J}_{\beta\alpha}).
\end{equation}
From Eq.~\ref{eq:dotpab-3}, we find that the rotation of a rod due to the flow induced by a tumbling neighbor scales as $L^3 \dot\gamma/\ln(L/4a)r^3$. Recall that non-tumbling rods (i.e., rods nearly aligned with the flow) induce only small to none flow disturbance. In semi-dilute concentrations, a given rod is mostly disturbed by the nearest tumbling neighbors for which the period of intercalation last for $\mathcal{O}(\dot\gamma^{-1})$. We may then model these interacting events as stochastic kicks in orientation, scaling as $\Theta\sim nL^3/\ln(L/4a)$, that occur with at a  frequency $\mathcal{O}(\rho)$, where $\rho$ is the rods mean tumbling frequency. \footnote{Here, we used the relation for the mean distance from the nearest neighbors, $\langle r\rangle^{-1/3}\sim n$, assuming a homogeneous spacial distribution of rods.} In this scenario, we end up a hydrodynamic driven mechanism for rotation diffusion characterized by the diffusion constant $D_h\sim \rho \Theta^2$. Note that this simplified model neglects any anisotropy of the diffusion dynamics, and is a parallel to the one described \citet{folgar1984orientation} but relying exclusive on hydrodynamic interactions and tumbling events. The arguments we use to build this model are much similar to the one discussed by \citet{rahnama1995effect}, mainly distinguishing by the fact that \citet{rahnama1995effect} considered the tumbling frequency to be the one from Jeffery's orbits while we allow for the self-sustained HI induced tumbling cascade, as elaborated in the main text. If we consider the high-Pe regime where $\rho\sim \Pe$, we are led to $D_h \sim \dot\gamma nL^3/\ln(L/4a)$ while \citet{rahnama1995effect} model suggests $D_h \sim \dot\gamma nL^3/[(L/2a)\ln^2(L/2a)]$. The simulation results shown in Fig. 2 in the main text aligns with the predictions of our model for varying aspect ratio.

\subsubsection{Estimating the diffusion dominated region}
\label{sec:highPe}
At high $\Pe$ numbers, orientation-averaged advection effects overcome diffusive ones, leading to a higher probability of rods aligned with the flow direction. However, there is a strong orientation dependence on the strength of advection. When the rods are highly aligned in the flow direction the effective role of advection is weaker. One can therefore define a region in which diffusion and advection are comparable. Hereafter, we refer to such region as the diffusion-dominated region (for reference, \citet{hinch1972effect} denoted this as the \textit{singular region}). In the diffusion-dominated region, the effective P\'eclet number $\Pe_{\textrm{eff}}$ is $\mathcal{O}(1)$. Following  \citet{hinch1972effect}, we define an appropriate angular scale for this region as $\delta$, whose magnitude we wish to find.

From Eq.~\ref{eq:rod-dynamics}, and using $\phi = \arctan{(p_y/p_z)}$ (the azimuthal angle in the $y\textrm{-}z$ plane) and $\theta = \arccos(p_x)$ (the orbital location in that azimuthal slice), we find the deterministic component of the rod orientation dynamics:
\begin{equation}
    \label{eq:dthetadt}
    \left\{\begin{matrix}  \dot\theta = \dot \gamma\sin^2\theta\sin\phi   \\
    \dot\phi = 0
\end{matrix}\right.  .
\end{equation}
For small values of $\theta$, the angular velocity scales as $\dot\gamma\delta^2$ and the respective advective time scales as $1/\delta\dot\gamma$. On the other hand, the characteristic diffusive time corresponding to the  diffusion-dominated region scales as $\delta^2/D$, where $D$ is the rotational diffusion constant. Since the effective P\'eclet number is the ratio of diffusive to advective times, we find that $\Pe_{\textrm{eff}}\sim D/\delta^3\dot\gamma$ and
\begin{equation}
    \label{eq:delta}
    \delta \sim (D/\dot\gamma)^{1/3}.    
\end{equation}

\subsection{Investigating the hydrodynamic cut-off distance}

In the present methodology, we implement a cut-off distance for pairwise hydrodynamic interactions based on the distance between the centers of mass of different rods $r^{\textrm{cm}}$. If the center-to-center distance $r^{\textrm{cm}}$ between two rods is smaller than the defined cut-off distance $r_c$, we account for the hydrodynamic interactions between all beads of the two rods using the Rotne-Prager-Yamakawa mobility tensor $\mathcal{M}_{ij}$ \cite{rotne1969, yamakawa1970}. For rods with  $r^{\textrm{cm}}>r_c$, the mobility tensor $\mathcal{M}_{ij}$ coupling the dynamics of beads of the two rods is set as zero (see methods). The use of the rod-based cut-off scheme instead of a bead-based scheme relies on the fact that the main non-hydrodynamic forces in the problem are internal to the rods and of elastic nature and, thus, reciprocal between neighboring beads. As such forces are the sources of the hydrodynamic disturbances, the reciprocity in the elastic interactions leads to force dipole hydrodynamics sources instead of monopole ones, the first having a faster decay than the latter (respectively, $1/r^2$ and $1/r$). Thus, interacting with a bead but not with its elastic counterpart introduces non-physical strong long-range disturbances. The rod-based cutoff scheme is a straightforward solution to address this issue.  This approach significantly reduces the computational costs of the simulations, allowing for the calculation of conclusive statistics for the studied concentrations. Yet, due to the long-range nature of the hydrodynamic interactions, the physical implications of $r_c$ must be evaluated carefully. Following, we investigate the role of $r_c$ at different concentrations. 

\begin{figure}[t]
    \centering
    \includegraphics[width=1\linewidth]{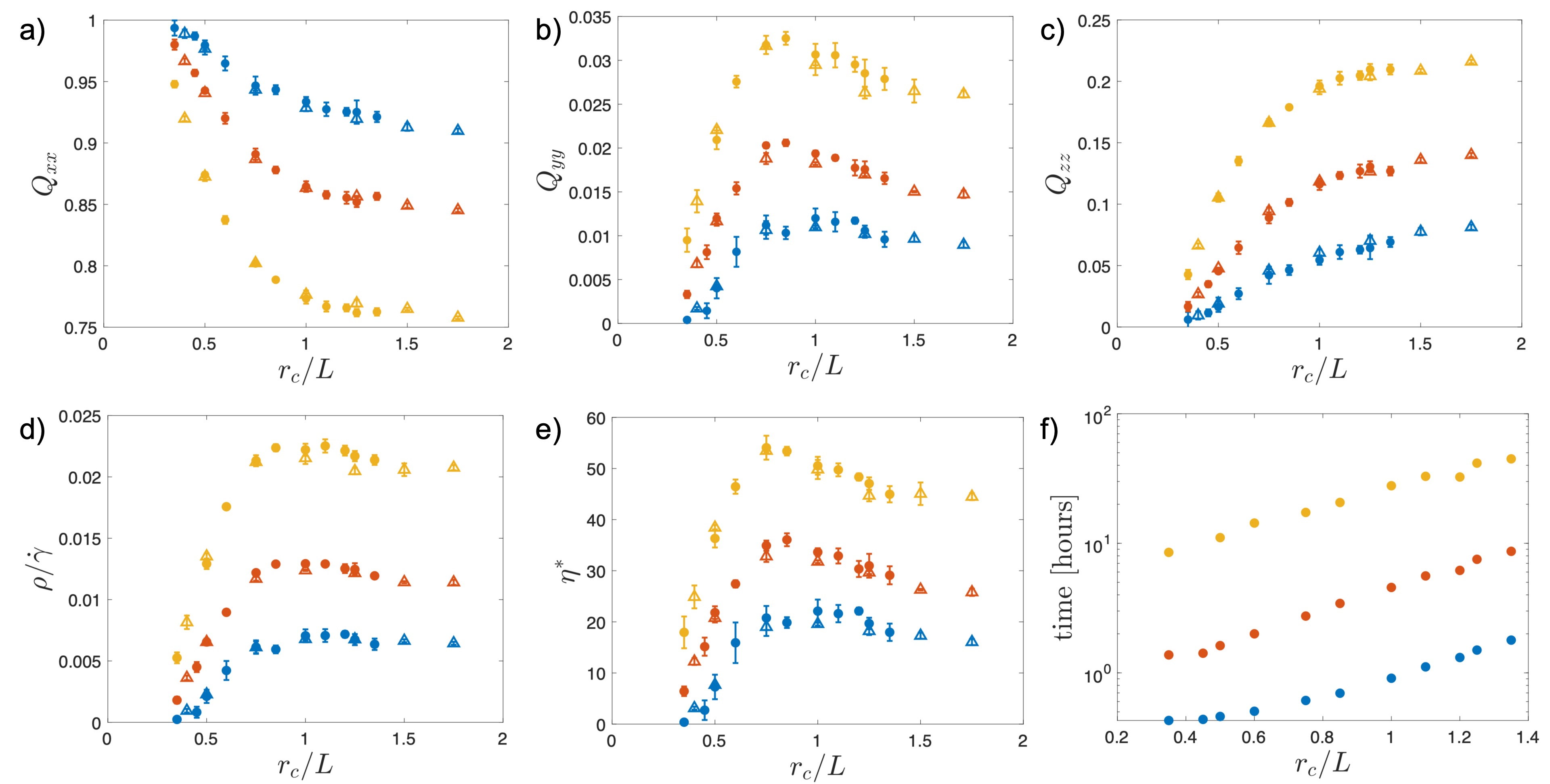}
    \caption{Results of the non-Brownian ($\kbT =0$) simulations as a function of the hydrodynamic cut-off length ($r_c/L$) for different concentration, $nL^3=1.11$ (blue symbols), $nL^3=2.22$ (red symbols) and $nL^3=4.44$ (yellow symbols). Panels a-f we present, respectively, $Q_{xx}$, $Q_{yy}$, $Q_{zz}$, $\rho/\dot\gamma$, $\eta^*$, and the clock simulating time. The simulations for the box of size $60^3$ are represented by the solid circles, and the simulations for the box of size $90^3$ are represented by the empty triangles.  }
    \label{fig:cutoff}
\end{figure}

We run non-Brownian simulations ($k_BT = 0$) for boxes of sizes $60^3$ (same used in the main work) and $90^3$ at the concentrations $nL^3=1.11$, 2.22, and 4.44. We analyze how the tumbling frequency  ($\rho/\dot\gamma$), the shear viscosity ($\eta^*$) and the components $Q_{xx}$, $Q_{yy}$ and $Q_{zz}$ of the second moment tensor $\textbf{Q}$, vary as functions of $r_c$. Results are compiled in Fig~\ref{fig:cutoff}. The simulations were run for a total strain of $5\times 10^3$, initiating from random orientations of the rods, i.e., isotropic. The only exception was the case $nL^3 = 4.44$ for the box of size $90^3$. For this specific case, we first run simulations for each value of $r_c/L$, starting from the isotropic configuration, for a total strain of $10^3$. The orientation distribution of the rods in the last few steps of the simulation is saved and randomly picked to define the rods' initial orientation in new simulations that are also run for more $10^3$ strains. Only the data corresponding to later runs are used in the analyses. For all cases in the box of size $60^3$, we run four to seven independent simulations. For the box of size $90^3$, three independent simulations are considered for each case. 

We observe that increasing $r_c/L$ leads to a gradual decrease in $Q_{xx}$ and increase in $Q_{zz}$, tending to plateau values for $r_c/L\gtrsim 1$. Alternatively, for $Q_{yy}$, we capture a non-monotonic behavior in which it initially increases, peaks at  $r_c/L \approx 0.8$, and slowly decreases to a plateau. The tumbling frequency ($\rho/\dot\gamma$) presents a similar behavior as $Q_{yy}$, although the peak and the later decrease are much weaker. These results bring interesting insights into the role of HI. The initial decrease in $Q_{xx}$ accompanied by the increase in $Q_{yy}$ and $Q_{zz}$ are direct and coupled results from the increasing $\rho/\dot\gamma$ described by the cascade tumbling effect elaborated in the main text. The more frequently the rods tumble, the less they align with the flow direction and, consequently, span in the other directions. However, for $r_c/L\gtrsim 1$ we note that $Q_{xx}$ mostly reaches a plateau while $Q_{zz}$ slowly increases at the detriment of $Q_{yy}$ decrease. This result highlights the slow decay of the hydrodynamic mechanism of pushing the rods' orientation towards the vorticity direction by inducing tumbling orbitals of smaller constant $C_b$. This slow decrease in $Q_{yy}$ for $r_c/L\gtrsim 1$ reflects in slow decrease of $\eta^*$ up to the plateau of both quantities. It is interesting to note that changing the box size from $60^3$ to $90^3$ led to no significant implications in the results, so we may consider our simulations free of box size effects and periodic boundary artifacts. Also, the behavior of the investigated parameters to $r_c/L$ is similar for all tested concentrations. 

In view of the high computational costs associated with increasing $r_c/L$ (see Fig.\ref{fig:cutoff}e), and the fact that there were no strong variations of the investigated measurements for $r_c/L\geq 1$ that could lead to distinct conclusions of the present investigation, we conducted our Brownian and non-Brownian investigations setting the cut-off length as $r_c/L=1$. It is important to mention that for the simulations with Brownian motion and HI, the highest computational cost is associated with the calculations to satisfy the thermal-fluctuation theorem, making the relative computational time of the Brownian simulations even more expensive than the ones presented in Fig.\ref{fig:cutoff}e.


\subsection{Average number of contacts per rod}
Here, we present a short investigation of the average number of contacts per rod for both the Brownian and non-Brownian cases. Although this is not the main focus of the present work, it still deserves some attention, since contacts and collisions in suspensions may be the source of diffusion mechanisms and may also lead to shear thickening responses depending on the friction coefficient and imposed shear rate \cite{rathee2020role}.

\begin{figure}[h]
    \centering
    \includegraphics[width=0.8\linewidth]{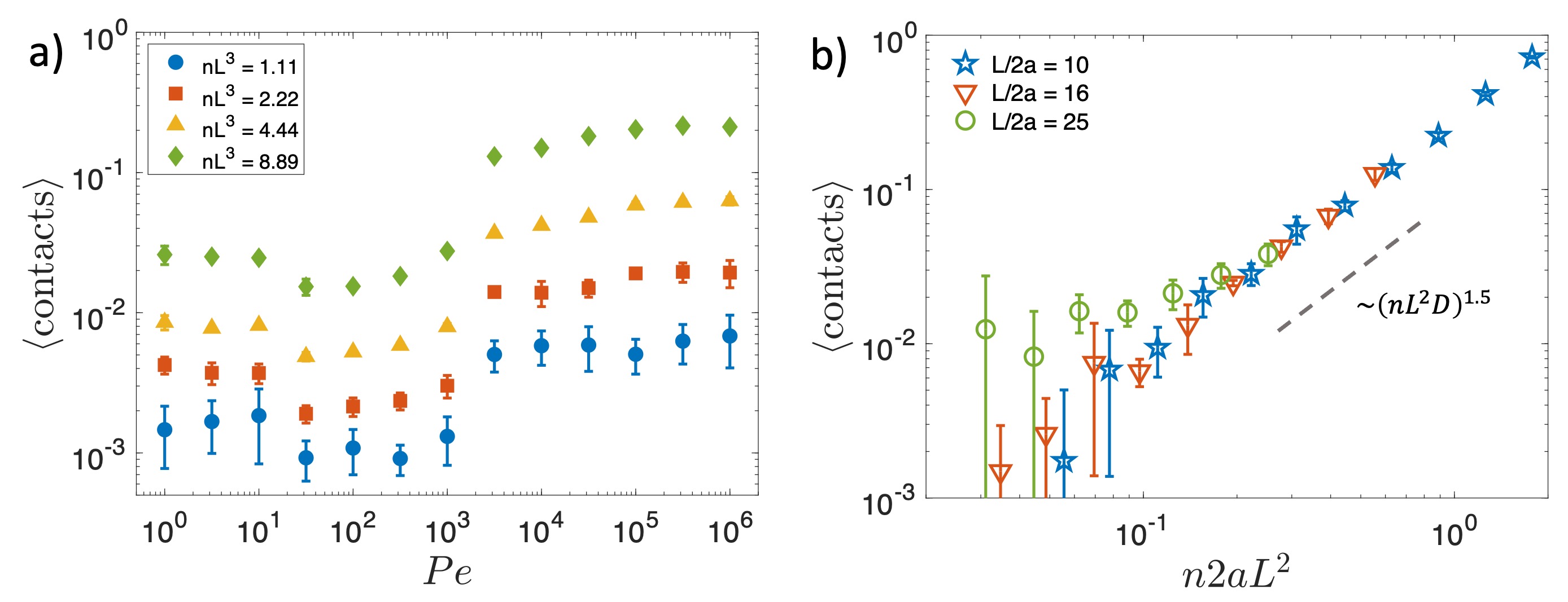}
    \caption{average number of contacts per rod  $\langle\text{contacts}\rangle$ for the Brownian (a) and non-Brownian cases (b) as function of $\Pe$ and $n2aL^2$, respectively.}
    \label{fig:contacts}
\end{figure}

Figure~\ref{fig:contacts} presents the average number of contacts per rod, $\langle\text{contacts}\rangle$, for the Brownian and non-Brownian cases. Although a given pair of rods may have multiple contact points due to the bead-spring discretization, this is counted as a single contact. For all Brownian cases, the aspect ratio is $L/2a = 10$. The results presented here correspond to the simulations shown in the main manuscript.

For the Brownian cases, $\langle\text{contacts}\rangle$ presents an initial slight decay with $\Pe$, in agreement with tube model theory, which states that flow-induced alignment of the rods increases the tube size and reduces the average number of contacts. Although this phenomenon is clearly observed when analyzing the number of contacts, its consequences on the order parameter $\mathcal{S}$ and the shear viscosity $\eta^*$ presented in the main text are not pronounced enough to be clearly identified. We believe this would require pushing the simulations to higher concentrations. At $\Pe \approx 10^3$, we observe a significant increase in $\langle\text{contacts}\rangle$, followed by a plateau. It is interesting to note that, although the system presents a higher degree of order at higher $\Pe$, the hydrodynamically induced cascade tumbling dynamics lead to the observed increase in $\langle\text{contacts}\rangle$. Despite the different absolute values, the response of $\langle\text{contacts}\rangle$ to variations in $\Pe$ behaves similarly for all concentrations.

For a system of rods at semi-dilute conditions, one expects $\langle\text{contacts}\rangle \sim nL^2D$, where $D = 2a$, based on excluded-volume considerations. Interestingly, in our non-Brownian simulations we find a stronger scaling, $\langle\text{contacts}\rangle \sim (nL^2D)^{1.5}$. We attribute this deviation to the non-equilibrium nature of the system under shear, in which, in addition to the geometric increase in encounter probability with concentration, the shear-driven cascade tumbling dynamics leads to the reduction of $\mathcal{S}$. This, in turn, increases the effective collision cross-subsection beyond the purely geometric estimate, yielding a superlinear dependence of $\langle\text{contacts}\rangle$ on $nL^2D$.

\subsection{Complementary data}

\begin{figure}[h]
    \centering
    \includegraphics[width=1\linewidth]{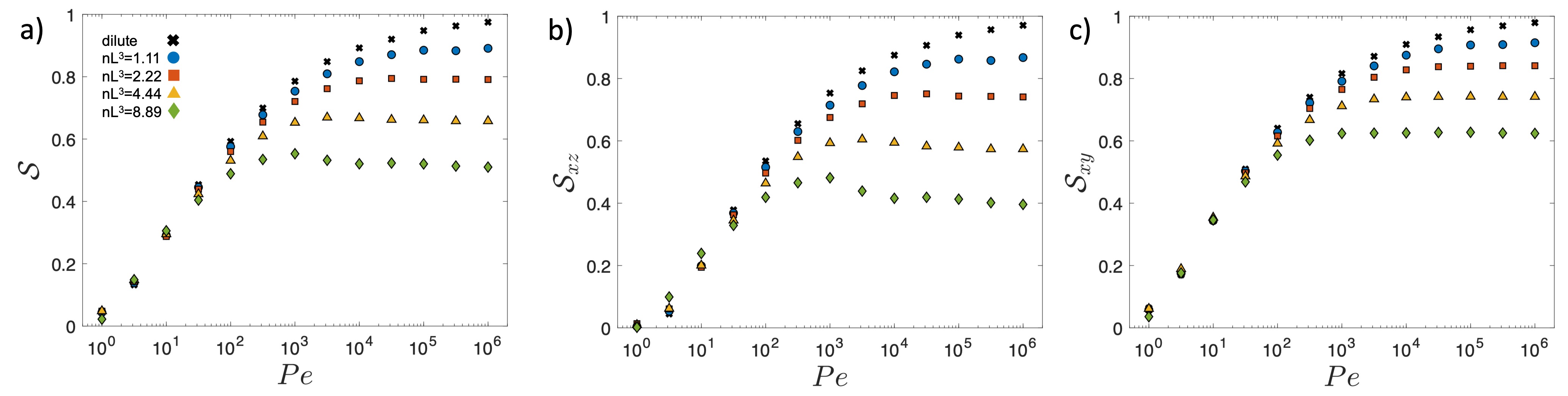}
    \caption{Order parameter $\mathcal{S}$, $\mathcal{S}_{xz}$, $\mathcal{S}_{xy}$ as a function of $\Pe$ at different $nL^3$. The presented results corresponds to the simulations of the main manuscript.   }
    \label{fig:S-vs-Pe}
\end{figure}

\newpage



\end{document}